\documentclass[10pt]{article}
\usepackage[english]{babel}
\usepackage[utf8]{inputenc}
\usepackage[T1]{fontenc} 
\usepackage{amsmath}
\usepackage{amssymb}
\usepackage{amsthm}
\usepackage{graphicx}
\usepackage{listings}
\usepackage{enumitem}
\usepackage[colorinlistoftodos]{todonotes}
\usepackage{float}
\usepackage[colorinlistoftodos]{todonotes}
\usepackage{mathtools}
\usepackage{setspace}
\usepackage{graphicx}
\usepackage{amsmath}

\usepackage{amssymb}

\numberwithin{equation}{section}
\newcommand{\eqa}{\begin{eqnarray}}
\newcommand{\eeqa}{\end{eqnarray}}
\newcommand{\beq}{\begin{equation}}
\newcommand{\eeq}{\end{equation}}

\begin{document}
\parskip 6pt
\hoffset -1.8cm

\title{Integrability,  geometry and wave solutions of some  Kairat  equations}
\author{Zh.  Myrzakulova$^{1}$\footnote{Email: zhrmyrzakulova@gmail.com},\, S. Manukure$^{2}$\footnote{Email: solomon.manukure@famu.edu}, \, R. Myrzakulov$^{1}$\footnote{Email: rmyrzakulova@gmail.com}, \,G. Nugmanova$^{1}$\footnote{Email: nugmanovagn@gmail.com}\\
\textsl{$^{1}$Ratbay Myrzakulov Eurasian International Centre for Theoretical} \\ {Physics, 010009, Astana, Kazakhstan}\\
\textsl{$^{2}$Department of Mathematics, Florida Agricultural and Mechanical} \\ {University, Tallahassee, USA}}
\maketitle

\begin{abstract} 
In this paper, we study some Kairat equations. The relation between the motion of  curves and Kairat equations is established. The geometrical equivalence between the Kairat-I equation and the Kairat-II equation is proved. We also proves that these equations is  gauge equivalent to each  other. Three types traveling wave solutions of the Kairat-II equation as well as its some integrals of motion are  found. The techniques used in this paper can be adopted to study  other integrable spin systems and nonlinear models. In particular, using these methods we study some Zhanbota equations.

{\bf Key words}:integrable systems, Lax pair, soliton,  Heisenberg ferromagnet  equation, Kairat-I  equation, gauge equivalence, Zhanbota-IV equation,  geometrical equivalence, Kairat-II  equation, travelling wave solutions, integrals of motion; Zhanbota-III equation. 
\end{abstract}


\section{Introduction}
The search for integrable nonlinear differential equations  (soliton equations)  has acquired enormous importance since the nonlinear
Kortweg-de Vries  equation, the nonlinear Schr\"{o}dinger equation and so on were  shown to be integrable. Such soliton  equations constitute a very special class of dynamical systems with many interesting properties.
The significance of constructiong  new  soliton equations, as well as investigating  their properties and
solutions, can not be overestimated. 

Integrable spin systems (ISS) are very important sector of integrable systems (see, e.g. \cite{2205.02073}-\cite{2202.00748} and references therein). They play an   important role in modern nonlinear physics and mathematics. The first representative  of such ISS  is the famous Heisenberg ferromagnet equation (HFE) \cite{L1}-\cite{Takhtajan1}
\begin{equation}
{\bf S}_{t}-{\bf S}\wedge {\bf S}_{xx}=0, \label{1.1}
\end{equation}
where ${\bf S}(x,t)=(S_{1},S_{2},S_{3})$ is the  spin vector, ${\bf S}^{2}=S_{1}^{2}+S_{2}^{2}+S_{3}^{2}=1$.
After the discovery of integrability of the HFE were proposed several ISS  in 1+1 dimensions. Here we can mention, for instance, the following  Landau-Lifshitz equation (LLE) \cite{2202.00748}
\begin{equation}
{\bf S}_{t}-{\bf S}\wedge {\bf S}_{xx} -{\bf S}\wedge J{\bf S}=0, \label{1.2}
\end{equation}
which is  the integrable anisotropic spin system. Here $J=diag(J_{1}, J_{2}, J_{3})$ and $J_{k}=const$.  Also there are some ISS  in 2+1 dimensions. The first example of the (2+1)-dimensional ISS is the Ishimori equation \cite{ishimori}
\begin{eqnarray}
{\bf S}_{t}-{\bf S}\wedge ({\bf S}_{xx}+{\bf S}_{yy})-u_{x}{\bf S}_{y}-u_{y}{\bf S}_{x}&=&0, \label{1.3} \\  
u_{xx}-\alpha^{2}u_{yy}+2\alpha^{2}{\bf S}\cdot({\bf S}_{x}\wedge{\bf S}_{y})&=&0, \label{1.4} 
\end{eqnarray}
where $u(x,y,t)$ is a real function (potential) and $\alpha=const$. Another example of the ISS in 2+1 dimensions is the Myrzakulov-I equation (M-IE) which has the form (see, e.g., \cite{s1}-\cite{z2} and references therein)
\begin{eqnarray}
{\bf S}_{t}-{\bf S}\wedge {\bf S}_{xy}-u{\bf S}_{x}&=&0,  \label{1.5}\\ 
u_{x}-2{\bf S}\cdot({\bf S}_{x}\wedge{\bf S}_{y})&=&0. \label{1.6} 
\end{eqnarray}
In the theory of integrable systems take place the so-called gauge and geometrical equivalences between integrable equations \cite{z1}-\cite{rm21}.  In particular, the HFE is equivalent to the nonlinear Schr\"{o}dinger equation \cite{L1}, \cite{ZT}, 
\begin{eqnarray}
iq_{t}+q_{xx}+2|q|^{2}q=0, \label{1.7} 
\end{eqnarray}
where $q(x,t)$ is a complex function. The gauge equivalent counterpart of the Ishimori equation (\ref{1.3})-(\ref{1.4})   is the following Davey-Stewartson equation 
\begin{eqnarray}
iq_{t}+q_{xx}+q_{yy}-vq&=&0, \label{1.8} \\
v_{xx}+\alpha^{2}v_{yy}-2(|q|^{2})_{xx}&=&0,\label{1.9} 
\end{eqnarray}
 where $v(x,t)$ is a real function (potential). At the same time, the geometrical and gauge equivalent of the M-IE (\ref{1.5})-(\ref{1.6}) is the following (2+1)-dimensional NLSE \cite{rm1}-\cite{rm21}
\begin{eqnarray}
iq_{t}+q_{xy}-vq&=&0, \label{1.10} \\
v_{x}+2(|q|^{2})_{y}&=&0.\label{1.11} 
\end{eqnarray}

In the present paper, we consider some  Kairat  equations  from a different point of view. In particular, we study their  integrability in the Lax sense, as well as their relations with the differential geometry of  curves and  equivalence aspects. Also we study some Zhanbota equations. 

The paper is organized as follows. In Section 2, we find a zero-curvature representation of the Kairat-I equation, with an essential (spectral) parameter. In Section 3, integrable motion of space curves induced by the Kairat-I equation is presented. The geometrical equivalent of the Kairat-I equation is considered in Section 4. The gauge equivalence  between these equations is established in Section 5.  The travelling wave solution of the Kairat-II equation is given in Section 6. The bilinear formulation of Kairat-II equation and its some conservation laws we discuss in Sections 7 and 8. In other sections we consider some other Kairat equations as well as the Zhanbota equations.  Section 14 is devoted to the conclusion.

\section{The Kairat-I equation}

The Kairat-I equation  has the following form
\begin{eqnarray}
{\bf S}_{xt}&=&\frac{u_{xt}}{u_{xxt}}(2u_{t}{\bf S}_{x}-{\bf S}_{t})+2(4u_{x}u_{t}-u_{xxt}){\bf S}+2u_{xt}{\bf S}\wedge{\bf S}_{x}, \label{2.1}\\
u_{x}&=&-\frac{1}{4}{\bf S}^{2}_{x}, \label{2.2}
\end{eqnarray}
where ${\bf S}(x,t)$ is the unit spin vector that is ${\bf S}=(S_{1},S_{2},S_{3})$ and ${\bf S}^{2}=S_{1}^{2}+S_{2}^{2}+S_{3}^{2}=1$, $u(x,t)$ is a real function (potential).  The Kairat-I equation is integrable. The corresponding Lax representation of the Kairat-I equation reads as
\begin{gather}
\Psi_x = U_{1} \Psi , \label{2.3} \\
\Psi_t = V_{1} \Psi , \label{2.4}
\end{gather}
where $U_{1}$ and $V_{1}$ are $2 \times 2$  matrix functions of $S$, $u$ and their finite-order derivatives with respect to $x$ and  $t$,  $\Psi (x,t)$ is a $2\times 2$ matrix function. The matrices $U_{1}$ and $V_{1}$ have the following  forms
\begin{eqnarray}
U_{1} &=& \frac{\lambda}{u_{xxt}}(2u_{t} S_{x}- S_{t}), \label{2.5}\\
V_{1}&=&\frac{\lambda}{1-4\lambda}\left(\frac{u_{t}+2u_{xxt}}{u_{xxt}}(2u_{t}S_{x}-S_{t})-u_{xt}S+\frac{1}{2}u_{x}[S,S_{x}]\right), \label{2.6}
\end{eqnarray}
where
\begin{gather}
S =
\begin{pmatrix}
S_{3} & S^{-} \\
S^{+} & -S_{3}
\end{pmatrix}
, \quad S^{2}=I, \quad S^{\pm}=S_{1}\pm iS_{2}, \quad S^{+}S^{-}+S_{3}^{2}=1. \label{2.7}
\end{gather}

The compatibility condition of the linear equations (\ref{2.3})-(\ref{2.4})
\begin{equation}
U_{1t}-V_{1x}+[U_{1},V_{1}]=0 \label{e8}
\end{equation}
gives the following set of nonlinear equations
\begin{eqnarray}
S_{xt}&=&\frac{u_{xt}}{u_{xxt}}(2u_{t}S_{x}-S_{t})+2(4u_{x}u_{t}-u_{xxt})S+u_{xt}[S,S_{x}], \label{2.9}\\
u_{x}&=&-\frac{1}{8}tr(S^{2}_{x}).\label{2.10}
\end{eqnarray}
It is the matrix form of the Kairat-I equation (\ref{2.1})-(\ref{2.2}).
\section{Integrable motion of space curves induced by the  Kairat-I equation}

Consider a space curve in the Euclidean space $R^{3}$. The corresponding Serret - Frenet equation and its temporal  counterpart are given by 
\begin{equation} 
\begin{pmatrix}\textbf{e}_1\\\textbf{e}_2\\\textbf{e}_3\end{pmatrix}_{x}=C \begin{pmatrix}\textbf{e}_1\\\textbf{e}_2\\\textbf{e}_3\end{pmatrix},  \\ \quad \begin{pmatrix}\textbf{e}_1\\\textbf{e}_2\\\textbf{e}_3\end{pmatrix}_{t}=D \begin{pmatrix}\textbf{e}_1\\\textbf{e}_2\\\textbf{e}_3\end{pmatrix}. \label{3.1}
\end{equation}
Here
\begin{equation} 
C=\begin{pmatrix}0&k_1&k_2 \\ -k_1&0&\tau \\ -k_2&-\tau&0 \end{pmatrix},  \\ \quad
D=\begin{pmatrix}0&\omega_3&\omega_2 \\ -\omega_3&0&\omega_1 \\ -\omega_2 &-\omega_1&0 \end{pmatrix}, \label{3.2}
\end{equation}
where  $\tau$ is the torsion, $k_1$ and  $k_2$ are the geodesic curvature and normal curvature of the curve, respectively. In Eq.(\ref{3.2}),  $\omega_j$   are some real functions depending on $x$ and $t$. The compatibility condition of the linear equations (\ref{3.1}) reads as
\begin{equation}  
C_t-D_x+[C,D]=0.
\end{equation}
In terms of elements this equation takes the form
\begin{eqnarray} 
\kappa_{1t}-\omega_{3x}-\kappa_2\omega_1+\tau\omega_2&=&0, \label{3.4}\\
\kappa_{2t}-\omega_{2x}+\kappa_1\omega_1-\tau\omega_3&=&0,  \label{3.5}\\
\tau_{t}-\omega_{1x}-\kappa_1\omega_2+\kappa_2\omega_3&=&0. \label{3.6}
\end{eqnarray}
In this paper, we assume that the  functions $\kappa_{j}, \tau, \omega_{j}$ have the following forms
\begin{equation} 
\kappa_1=0, \quad
\kappa_2=u_{x}-\lambda-1,  \quad
\tau=i(u_{x}-\lambda +1), \label{3.7}
\end{equation} 
and
\begin{eqnarray} 
\omega_1&=&-\frac{i}{1-4\lambda}[u_{xxt}+2u_{t}(u_{x}-\lambda+1)],  \\  \label{3.8}
\omega_2&=&\frac{1}{1-4\lambda}[2u_{t}(u_{x}-\lambda -1)-u_{xxt}], \\  \label{3.9}
\omega_3&=&\frac{2i}{1-4\lambda}u_{xt}. \label{3.10}
\end{eqnarray}
In this case, the unit vector ${\bf e}_{3}$ satisfies the following equation
\begin{eqnarray}
{\bf e}_{3xt}&=&\frac{u_{xt}}{u_{xxt}}(2u_{t}{\bf e}_{3x}-{\bf e}_{3t})+2(4u_{x}u_{t}-u_{xxt}){\bf e}_{3}+2u_{xt}{\bf e}_{3}\wedge{\bf e}_{3x}, \label{3.11}\\
u_{x}&=&-\frac{1}{4}{\bf e}^{2}_{3x}. \label{3.12}
\end{eqnarray}
It is nothing but the Kairat-I equation (\ref{2.1})-(\ref{2.2}) written in terms of the unit vector function ${\bf e}_{3}(x,t)$ and the potential function  $u(x,t)$. This result proves that the motion of the space curve with expressions (\ref{3.7})-(\ref{3.10}) that is induced by the Kairat-I equation (\ref{2.1})-(\ref{2.2}) is integrable.

\section{Geometrical equivalence. The Kairat-II equation}

Let us now we find the so-called geometrical equivalent of the Kairat-I equation (\ref{2.1})-(\ref{2.2}). Substituting the expressions for the functions $\kappa_{j}, \tau, \omega_{j}$ (\ref{3.7})-(\ref{3.10}) into the set of equations (\ref{3.4})-(\ref{3.6}) we obtain the following equation for the function $u(x,t)$:
\begin{eqnarray}
u_{xt}-2u_{t}u_{xx}-4u_{x}u_{xt}+u_{xxxt}=0. \label{4.1}
\end{eqnarray}
It is the Kairat-II equation. So we have proved that the Kairat-I equation (\ref{2.1})-(\ref{2.2}) and the Kairat-II equation (\ref{4.1}) geometrically equivalent to each  other. Since the Kairat-I equation (\ref{2.1})-(\ref{2.2}) is integrable, its equivalent counterpart,  the  Kairat-II equation (\ref{4.1}) is also integrable. The  Lax representation of the Kairat-II equation  (4.1) is given by
\begin{eqnarray}
\Phi_{x}&=&U_{2}\Phi, \label{4.2}\\
\Phi_{t}&=&V_{2}\Phi, \label{4.3}
\end{eqnarray}
where
\begin{eqnarray}
U_{2}&=&\lambda \Sigma+Q, \label{4.4}\\
V_{2}&=&\frac{1}{1-4\lambda}(\lambda B_{1}+B_{0}). \label{4.5}
\end{eqnarray}
Here
\begin{eqnarray}
\Sigma&=&\begin{pmatrix}0&0 \\ -1&0  \end{pmatrix},   \quad
Q=\begin{pmatrix}0& 1 \\ u_{x} & 0
\end{pmatrix}, \quad B_{1}=\begin{pmatrix}0& 0 \\ -2u_{t} & 0
\end{pmatrix}=2u_{t}\Sigma, \\
B_{0}&=&\begin{pmatrix}-u_{xt}& 2u_{t} \\2u_{x}u_{t}-u_{xxt} & u_{xt}
\end{pmatrix}=-u_{xt}\sigma_{3}+2u_{t}Q+u_{xxt}\Sigma. 
\end{eqnarray}
The compatibility condition of the set of linear equations (\ref{4.2})-(\ref{4.3}) 
\begin{eqnarray}
U_{2t}-V_{2x}+[U_{2},V_{2}]=0 \label{4.8}
\end{eqnarray}
gives the Kairat-II equation (4.1). 
\section{Gauge equivalence}
It is well-known that between some integrable equations takes place the so-called gauge equivalence. In our case, the Kairat-I equation (\ref{2.1})-(\ref{2.2}) and the Kairat-II equation (\ref{4.1}) are gauge equivalent to each  other. To prove this statement, let us consider the gauge transformation
\begin{eqnarray}
\Psi=g^{-1}\Phi, \label{5.1}
\end{eqnarray}
where $g(x,t)=\Phi(x,t,\lambda)|_{\lambda=0}$. Then the Lax pairs $U_{1}, V_{1}$ (\ref{2.5})-(\ref{2.6}) and $U_{2}, V_{2}$ (\ref{4.4})-(\ref{4.5}) are  related by the following transformations
\begin{eqnarray}
U_{1}&=&g^{-1}U_{2}g-g^{-1}g_{x},\\ 
V_{1}&=&g^{-1}V_{2}g-g^{-1}g_{t}. 
\end{eqnarray}
This results proves the gauge equivalence between the Kairat-I equation (\ref{2.1})-(\ref{2.1}) and the Kairat-II equation (\ref{4.1}).  From this equivalence follow some very important relations between solutions of these K-I and K-II equations. Let us present some of them:
\begin{eqnarray}
u_{t}=\frac{tr(S_{x}S_{t})}{2tr(S_{x}^{2})}, \quad u=-\frac{1}{8}\int tr(S_{x}^{2})dx, \quad u_{xt}=-\frac{1}{8}[tr(S_{x}^{2})]_{t}, \quad tr(S_{x}^{2})=-8u_{x}. \label{5.4}
\end{eqnarray}
\section{Travelling solutions}
In this section we are going to present exact three travelling wave solutions of the Kairat-II equation (4.1): elliptical, soliton and rational solutions. Let us find the travelling wave solution of the K-IIE (4.1). To do that instead of the independent variables $(x,t)$,  let us introduce the new independent variable $y=a x +bt+c$, where $a,b,c$ are real constants. So we have $u(x,t)=u(y)$ and 
\begin{eqnarray}
u^{'}=\frac{\partial u}{\partial y}, \quad u_{t}=bu^{'}, \quad u_{x}=au^{'}, \quad u_{xx}=a^{2}u^{''}, \quad u_{xt}=abu^{''}, \quad 
u_{xxxt}=a^{3}bu^{''''}.
\end{eqnarray}
After some simple algebraic manipulations of the following forms of the K-II equation (4.1), we obtain:
\begin{eqnarray}
abu^{''}-6a^{2}bu^{'}u^{''}+a^{3}bu^{''''}=0,
\end{eqnarray}
\begin{eqnarray}
u^{''}-3a(u^{'2})^{'}+a^{2}u^{''''}=0,
\end{eqnarray}
\begin{eqnarray}
u^{'}-3au^{'2}+a^{2}u^{'''}+n=0,
\end{eqnarray}
\begin{eqnarray}
2u^{'}u^{''}-6au^{'2}u^{''}+2a^{2}u^{''}u^{'''}+2nu^{''}=0,
\end{eqnarray}
\begin{eqnarray}
(u^{'2})^{'}-2a(u^{'3})^{'}+a^{2}(u^{''2})^{'}+2nu^{''}=0.
\end{eqnarray}
Finally we get  the following ordinary differential equation
\begin{eqnarray}
u^{'2}-2au^{'3}+a^{2}u^{''2}+2nu^{'}+m=0,
\end{eqnarray}
where $m$ and $n$ are  real integration constants. Let $u^{'}(y)=f(y)$, where $f(y)$ is a real function.  Then the last equation takes the form
\begin{eqnarray}
a^{2}f^{'2}-2af^{3}+f^{2}+2nf+m=0.
\end{eqnarray}
Our next step is $f(\zeta)=\mu+kh(\zeta)$, where $\mu,k$ are some real constants and $h(y)$ is a real function of $y$.  Then step by step we obtain following equations for the functions $h(y)$:
\begin{eqnarray}
a^{2}k^{2}h^{'2}-2a(\mu+kh)^{3}+(\mu+kh)^{2}+2n(\mu+kh)+m=0,
\end{eqnarray}
\begin{eqnarray}
a^{2}k^{2}h^{'2}-2a(\mu^{3}+3\mu^{2}kh+3\mu k^{2}h^{2}+k^{3}h^{3})+(\mu^{2}+2\mu kh+k^{2}h^{2})+2n(\mu+kh)+m=0,
\end{eqnarray}
\begin{eqnarray}
a^{2}k^{2}h^{'2}=2ak^{3}h^{3}+(6a\mu k^{2}-k^{2})h^{2}+(6a\mu^{2}k-2\mu k-2nk)h+(2a\mu^{3}-\mu^{2}-2n\mu-m)=0,
\end{eqnarray}
\begin{eqnarray}
a^{2}k^{2}h^{'2}=2ak^{3}h^{3}+k^{2}(6a\mu-1)h^{2}+k(6a\mu^{2}-2\mu-2n)h+(2a\mu^{3}-\mu^{2}-2n\mu-m)=0, \label{5.12}
\end{eqnarray}
\begin{eqnarray}
h^{'2}=\frac{2k}{a}h^{3}+a^{-2}(6a\mu-1)h^{2}+\frac{(6a\mu^{2}-2\mu-2n)}{a^{2}k}h+\frac{(2a\mu^{3}-\mu^{2}-2n\mu-m)}{a^{2}k^{2}}=0. \label{6.13}
\end{eqnarray}
\subsection{Elliptic solution}
Now we introduce two constants $g_{2}$ and $g_{3}$ as: 
\begin{eqnarray}
 g_2=-\frac{(6a\mu^{2}-2\mu -2n)}{a^{2}k}, \quad g_3=-\frac{(2a\mu^{3}-\mu^{2}-2n\mu-m)}{a^{2}k^{2}}.
\end{eqnarray}
Then the unknown integration constants $n,m$ have the forms:
\begin{eqnarray}
n=a^{2}kg_{2}+3a\mu^{2}-\mu, \quad m=g_{3}a^{2}k^{2}+2a\mu^{3}-\mu^{2}-2n\mu.
\end{eqnarray}
Now we assume that  constants $k$ and $\mu$ have the forms
\begin{eqnarray}
k=2a, \quad \mu=\frac{1}{6a}.
\end{eqnarray}
Then the equation (\ref{6.13}) takes the form
\begin{eqnarray}
h'^2(y) = 4h^3(y)-g_2h-g_3, \label{6.17}
\end{eqnarray}
which has the same form with the equation for the Weierstrass function $\wp$:\begin{eqnarray}
\wp'^2(y) = 4\wp ^3(y)-g_2\wp(y)-g_3.
\end{eqnarray}
Thus we obtain  
\begin{eqnarray}
h(y)=\wp(y;\Lambda)=-\zeta^{'}(y;\Lambda),
\end{eqnarray}
where $\zeta(y; \Lambda)$ is the Weierstrass zeta function.  Hence we obtain
\begin{eqnarray}
 u^{'}=f=\mu-2a\zeta^{'}(y).
\end{eqnarray}
 Thus finally we have 
\begin{eqnarray}
u(x,t)=u(y)=\nu+\mu y-2a\zeta(y),
\end{eqnarray}
where $\nu$ is a real integration constant. It is the desired exact  travelling wave solution of the Kairat-II equation (4.1) in the elliptic function form.
\subsection{Soliton solution}
In this subsection let us find the soliton solution of the Kairat-II equation (\ref{4.1}). To do that we assume that 
\begin{eqnarray}
h(y)=\alpha \sec^{2}(y)+\beta=\frac{\alpha}{\cosh^{2}(y)}+\beta, \label{5.22}
\end{eqnarray}
where $\alpha, \beta$ are some real constants. We have
\begin{eqnarray}
h^{'}(y)=-\frac{2\alpha\sinh(y)}{\cosh^{3}(y)},\quad h^{'2}=4\alpha^{2}\left(\frac{1}{\cosh^{4}(y)}-\frac{1}{\cosh^{6}(y)}\right).\label{5.23}
\end{eqnarray}
Substituting these expressions into (\ref{6.13}) we obtain
\begin{eqnarray}
\cosh^{-6}(y)&:& \alpha=-1, \\
 \cosh^{-4}(y)&:& \beta=\frac{1}{3}, \\
\cosh^{-2}(y)&:& g_{2}=\frac{4}{3},\\
\cosh^{-0}(y)&:&  g_{3}=\frac{4}{27}.
\end{eqnarray}
Thus we obtain
\begin{eqnarray}
h(y)=\frac{1}{3}-\frac{1}{\cosh^{2}(y)}, \label{5.12}
\end{eqnarray}
or
\begin{eqnarray}
f(y)=\frac{1}{6a}+2ah(y)=\frac{1}{6a}+2a\left(\frac{1}{3}-\frac{1}{\cosh^{2}(y)}\right), \label{5.12}
\end{eqnarray}
\begin{eqnarray}
u(y)=\int f(y)dy=\rho+\left(\frac{1}{6a}+\frac{2a}{3}\right)y-2a\tanh(y), \label{5.12}
\end{eqnarray}
Thus the simplest soliton  solution of the Kairat-II equation has the following form
\begin{eqnarray}
u(x,t)=u(y)=\rho+\left(\frac{1}{6a}+\frac{2a}{3}\right)y-2a\tanh(y), \label{5.12}
\end{eqnarray}
where $\rho=const$.
\subsection{Rational solution}
We now want to present the rational solution of the Kairat-II equation (\ref{4.1}). We look for such solutions in the form
\begin{eqnarray}
h(y)=\frac{\delta}{y^{2}}+\epsilon, \label{6.32}
\end{eqnarray}
where $\delta$ and $\epsilon$ are real constants. This expression satisfies the equation (\ref{6.17}) if
\begin{eqnarray}
\delta=1, \quad \epsilon=g_{2}=g_{3}=0. \label{6.33}
\end{eqnarray}
Hence and from (\ref{6.13}) we get
\begin{eqnarray}
n=3a\mu^{2}, \quad m=-4a\mu^{3}-\mu^{2}, \quad \mu=\frac{1}{6a}. \label{6.34}
\end{eqnarray}
Finally we obtain the following rational solution of the Kairat-II equation (4.1) as
\begin{eqnarray}
u(x,t)=u(y)=\eta +\mu y-\frac{k}{y}, \label{6.35}
\end{eqnarray}
where $\eta=const$.

\section{Bilinear formulation}
In this section, we formulate bilinear equation for the  Kairat-II equation \eqref{4.1}. Under the transformation 
$$u=-2(\ln f)_x$$
equation \eqref{4.1} can be transformed into
\begin{eqnarray}\label{7.1}
	D_x\left[(D_x^3D_t+D_xD_t)f\cdot f\right]\cdot f^2+\frac{1}{3}D_t[D_x^4f\cdot f]\cdot f^2-\frac{1}{3}D_x[D_x^3D_t f\cdot f]\cdot f^2=0
\end{eqnarray}
where the $D$-operators are the Hirota  bilinear operators defined by \cite{Hirota}
\begin{equation}
	D_x(f\cdot g)=\left(\frac{\partial}{\partial x}-\frac{\partial}{\partial x'}\right)f(x)\cdot g(x')\bigg|_{x'=x}
\end{equation}
or more generally,

\begin{equation*}
	D_{x}^{m}D_{t}^{n} (f\cdot g)=\left(\frac{\partial}{\partial x}-\frac{\partial}{\partial x'}\right)^m\left(\frac{\partial}{\partial t}-\frac{\partial}{\partial t'}\right)^nf(x,t)\cdot g(x',t')\bigg|_{x'=x,t'=t}.
\end{equation*}
The above equation \eqref{7.1} is not bilinear. Observe that all the terms in \eqref{7.1} are of the form $D_x[\cdots]\cdot f^2$, except the middle term. We introduce a new independent variable $\tau$ and impose the auxiliary condition
\begin{equation}
	D_x(D_x^3+D_{\tau})f\cdot f=0
\end{equation} 
Now, using the identity
\begin{equation}
	D_t(D_xD_{\tau})f\cdot f=D_x(D_tD_{\tau})f\cdot f
\end{equation}
we obtain the following bilinear form for the Kirat-II equation:
\begin{equation}
	\begin{cases}
	D_x(D_x^3+D_{\tau})f\cdot f=0\\
	\displaystyle
	\left(D_x(D_t+D_x^2D_t)-\frac{1}{3}D_t(D_{\tau}+D_x^3)\right)f\cdot f=0
	\end{cases}
\end{equation}

\section{Conservation  laws (integrals of motion)}
As integrable system, the Kairat equation (4.1) admits the infinite number conservation  laws. In this section we give some equations which allow to construct  these conservation  laws of the Kairat-II equation (4.1). To find   the conservation laws, let us introduce two functions as
\begin{eqnarray}
\Gamma_{1}=\frac{\phi_{1}}{\phi_{2}}, \quad \Gamma_{2}=\frac{\phi_{2}}{\phi_{1}}. \label{8.1}
\end{eqnarray}
These functions satisfy the following set of nonlinear equations
\begin{eqnarray}
\Gamma_{1x}&=&1-(u_{x}-\lambda)\Gamma_{1},   \\
\Gamma_{1t}&=&\frac{1}{1-4\lambda}[2u_{t}-2u_{xt}\Gamma_{1}-(2u_{x}u_{t}-u_{xxt}-2u_{xx}\lambda)\Gamma_{1}^{2}], \\
\Gamma_{2x}&=&u_{x}-\lambda -\Gamma_{2}^{2},   \\
\Gamma_{2t}&=&\frac{1}{1-4\lambda}[(-2\lambda u_{t}+2u_{x}u_{t}-u_{xxt})+2u_{xt}\Gamma_{2}-2u_{t}\Gamma_{2}^{2}].
\end{eqnarray}
Using these equations we can construct the infinite number of conservation  laws.  As an example, in this section we give in explicit forms  two of these conservation laws (integrals of motion). Their equations have the forms:
\begin{eqnarray}
I_{1t}&=&J_{1x},   \\
I_{2t}&=&J_{2x},
\end{eqnarray}
where
\begin{eqnarray}
I_{1}&=&(u_{x})^{2},   \\
I_{2}&=&u_{x}-2(u_{x})^{2}+u_{xxx}, \\
J_{1}&=&u_{t}-2u_{x}u_{t}+u_{xxt},   \\
J_{2}&=&2u_{x}u_{t}.
\end{eqnarray}

 
\section{Kairat-III equation}
\subsection{Kairat-III equation}
The  Kairat-III equation has the form
\begin{eqnarray}
i(1-iK_{1})S_t+\frac{1}{2}[S,N_1S]+K_2S_x&=& 0, \\
 N_2u-\frac{\alpha^{2}}{4i}tr(S[S_{t},S_x])&=&0, 
\end{eqnarray}
where 
\begin{eqnarray}
N_1&=& \alpha ^2\frac{\partial ^2}{\partial t^2}-2\alpha (b-a)\frac{\partial^2}
   {\partial x \partial t}+(a^2-2ab-b)\frac{\partial^2}{\partial x^2},\\
N_2&=&\alpha^2\frac{\partial^2}{\partial t^2} -\alpha(2a+1)\frac{\partial^2}
   {\partial x \partial t}+a(a+1)\frac{\partial^2}{\partial x^2},\\
	K_1&=&2i\{(2ab+a+b)u_x-(2b+1)\alpha u_{t}\}, \\
K_2&=&2i\{(2ab+a+b)u_{t}-\alpha^{-1}(2a^2b+a^2+2ab+b)u_x\},
\end{eqnarray}
and  $\alpha, b, a$=  consts. This Kairat-III equation is integrable. 
\subsection{Gauge equivalent}
To prove integrability of the Kairat-III  equation (9.1)-(9.2), let us we consider the following transformation (representation) for the spin matrix function $S(x,t)$:
\begin{eqnarray}
S=\Phi^{-1}\sigma_{3}\Phi,
\end{eqnarray}
where $\Phi(x,t)$ is some matrix function. In this section, we assume that $\Phi$ satisfies the following equation
\begin{eqnarray}
\alpha D_{2}\Phi_{xx}=D_{1}\Phi_{x}+D_{0}\Phi,
\end{eqnarray}
where $D_{j}$ are some matricies. In this paper, we assume that these matrices have the following forms:
\begin{eqnarray}
D_{2}=\begin{pmatrix}b+1&0 \\ 0&b  \end{pmatrix},   \quad
D_{1}=i\begin{pmatrix}-a-1& i\alpha q \\ i\alpha r & -a
\end{pmatrix}, \quad D_{0}=\begin{pmatrix}\alpha c_{11}& \alpha c_{12}-q \\ 
\alpha c_{21}-r & \alpha c_{22}
\end{pmatrix}. 
\end{eqnarray}
Here $c_{ij}$ satisfy the following equations
\begin{eqnarray}
\alpha c_{11t}-(a+1)c_{11x}&=&qc_{21}-rc_{12}-iqr_{x}, \\
\alpha c_{22t}-ac_{22x}&=&rc_{12}-qc_{21}-irq_{x},\\
c_{12}&=&i\alpha q_{t}+i(2b-a+1)q_{x}, \\
c_{21}&=&-ir_{t}+i(a-2b)r_{x}.
\end{eqnarray}
Now substituting Eq.(9.7) into the Kairat-III equation  (9.1)-(9.2), after some algebra,  we obtain that the matrix function $\Phi$ obeys the following equation
\begin{eqnarray}
\alpha \Phi_{t}=B_{1}\Phi_{x}+B_{0}\Phi, 
\end{eqnarray}
where
\begin{eqnarray}
B_{1}=\begin{pmatrix}a+1&0 \\ 0&a  \end{pmatrix},   \quad
B_{0}=\begin{pmatrix}0& q \\ r & 0
\end{pmatrix}. 
\end{eqnarray}
Thus we have the following two linear  equations for the matrix function $\Phi(x,t)$:
\begin{eqnarray}
\alpha D_{2}\Phi_{xx}&=&D_{1}\Phi_{x}+D_{0}\Phi,\\
\alpha \Phi_{t}&=&B_{1}\Phi_{x}+B_{0}\Phi.
\end{eqnarray}
The compatibility condition of these equations gives the following Kairat-IV equation:
\begin{eqnarray}
iq_{t}+N_{1}q+vq&=&0,\\
ir_{t}-N_{1}r-vr&=&0,\\
N_{2}v+2N_{1}(rq)&=&0,
\end{eqnarray}
where $r=\epsilon \bar{q} \quad (\epsilon=\pm 1)$, $q(x,t)$ is a complex-valued function representing the wave profile. In this system, the first  term is the linear temporal evolution, the second term represents the dispersion term, while the third term represents the nonlinearity term. This  Kairat-IV equation (9.18)-(9.20) is the gauge equivalent counterpart of the Kairat-III equation (9.1)-(9.2). Since the Kairat-IV equation (9.18)-(9.20) is integrable with the Lax representation (9.16)-(9.17), its gauge equivalent counterpart - the Kairat-III equation (9.1)-(9.2) is also integrable. As integrable equations the Kairat-III equation and the Kairat-IV equation admit the N-soliton solutions, the infinite number of integrals ob motion and so on. 

Finally we present the more gemeral form of the K-IVE, namely, the following one
\begin{eqnarray}
iq_{t}+N_{3}q+vq&=&0,\\
ir_{t}-N_{3}r-vr&=&0,\\
N_{4}v+2N_{3}(rq)&=&0,
\end{eqnarray}
with
\begin{eqnarray}
N_{3}&=& a_{1}\frac{\partial ^2}{\partial t^2}+a_{2}\frac{\partial^2}
   {\partial x \partial t}+a_{3}\frac{\partial^2}{\partial x^2},\\
N_{4}&=&b_{1}\frac{\partial^2}{\partial t^2} +b_{2}\frac{\partial^2}
   {\partial x \partial t}+b_{3}\frac{\partial^2}{\partial x^2},
\end{eqnarray}
where $a_{j}$ and $b_{j}$ are some real constants.

\subsection{Hirota bilinear form of the Kairat-IV equation}

To construct the soliton solutions we can use the Hirota bilinear method. Let us find the Hirota bilinear form of the Kairat-IV equation (9.18)-(9.20). To do that we introduce the following representation for the function $q(x,t)$:
\begin{eqnarray}
q=\frac{g}{f}, \quad v=-2N_{1}\ln f,
\end{eqnarray}
where $q(x,t)$ is a complex function and $f(x,t)$ is a real function. Substituting these formulas to Eqs. (9.18)-(9.2)) we get
\begin{eqnarray}
\left(iD_{t}+\alpha^{2}D_{t}^{2}-2\alpha (b-a)D_{t}D_{x}+(a^{2}-2ab-b)D_{x}^{2}\right)g\cdot f&=&0,\\
\left(\alpha^{2}D_{t}^{2}-\alpha (2a+1)D_{t}D_{x}+a(a+1)D_{x}^{2}\right)f\cdot f- rq&=&0.
\end{eqnarray}
It is the Hirota bilinear form of the Kairat-IV equation. Using this bilinear form we can construct  different soliton solutions of the Kairat-IV equation. 
\section{Kairat-V  equation}
In this section, we consider the following Kairat-V equation
\begin{eqnarray}
S_{xt}-\frac{2\lambda f_{xxx}}{f_{xx}}\left(\frac{2\lambda+1}{2\lambda}S_{t}+2fS_{x}\right)-\frac{4\lambda(f_{xx}-4\lambda mf)}{2\lambda+1}S+\frac{2\lambda f_{x}}{2\lambda+1}[S,S_{x}]&=&0,\\
f_{xx}-\frac{2\lambda+1}{8\lambda}tr(S[S_{x},S_{t}])&=&0,
\end{eqnarray}
where $f(x,t)$ is a real function, $\lambda$ is a complex constant  and 
\begin{eqnarray}
m=-\frac{1}{8\lambda}tr(S_{x}^{2}).
\end{eqnarray}
It is interesting to note that the Kairat-V equation (10.1)-(10.2) is integrable. To prove this statement we consider the following representation for the spin matrix $S(x,t)$:
\begin{eqnarray}
S=\Phi^{-1}\sigma_{3}\Phi,
\end{eqnarray}
where $\Phi(x,t)$ is some matrix function. Plugging this representation into the Kairat-V equation (10.1)-(10.2) we obtain that the function $\Phi$ satisfies the following equations
\begin{eqnarray}
\Phi_{x}&=&U_{3}\Phi, \\
\Phi_{t}&=&V_{3}\Phi, 
\end{eqnarray}
where $\Phi=(\phi_{1},\phi_{2})^{T}$ and 
\begin{eqnarray}
U_{3}&=&\begin{pmatrix}0&1\\ \lambda m&0  \end{pmatrix},   \\
V_{3}&=&\frac{\lambda}{2\lambda+1}\begin{pmatrix}f_{x}&-2f \\ f_{xx}-2\lambda mf & -f_{x}
\end{pmatrix}=\frac{\lambda}{2\lambda+1}B_{3}. 
\end{eqnarray}
The compatibility condition of the equations (10.5)-(10.6)  
\begin{eqnarray}
U_{3t}-V_{3x}+[U_{3},V_{3}]=0
\end{eqnarray}
is equivalent to the following Kairat-VI equation 
\begin{eqnarray}
u_{xxt}+2u_{xx}f_{x}+u_{xxx}f&=&0,\\
f_{x}+u_{t}+c_{1}x+c_{2}&=&0,
\end{eqnarray}
where $c_{j}$ are some real constants. We can rewrite this Kairat-VI equation as
\begin{eqnarray}
m_{t}+2mf_{x}+m_{x}f&=&0,\\
f_{xxx}-m_{t}&=&0,
\end{eqnarray}
where $m=-u_{xx}$. The matrix  Lax representation (10.5)-(10.6) can be rewritten in scalar form as
\begin{eqnarray}
\phi_{1xx}&=&\lambda m\phi_{1}, \\
\phi_{1t}&=&\frac{\lambda}{2\lambda+1}(f_{x}\phi_{1}-2f\phi_{1x}).
\end{eqnarray}

\section{Integrable motion of two interationg curves: K-VIIE and K-VIIIE}

Consider two interacting   space curves in the Euclidean space $R^{3}$. The corresponding two Serret - Frenet equations and their  temporal  counterparts are given by 
\begin{equation} 
\begin{pmatrix}\textbf{e}_1\\\textbf{e}_2\\\textbf{e}_3\end{pmatrix}_{x}=C_{1} \begin{pmatrix}\textbf{e}_1\\\textbf{e}_2\\\textbf{e}_3\end{pmatrix},  \\ \quad \begin{pmatrix}\textbf{e}_1\\\textbf{e}_2\\\textbf{e}_3\end{pmatrix}_{t}=D_{1} \begin{pmatrix}\textbf{e}_1\\\textbf{e}_2\\\textbf{e}_3\end{pmatrix}. \label{11.1}
\end{equation}
and
\begin{equation} 
\begin{pmatrix}\textbf{l}_1\\\textbf{l}_2\\\textbf{l}_3\end{pmatrix}_{x}=C_{2} \begin{pmatrix}\textbf{l}_1\\\textbf{l}_2\\\textbf{l}_3\end{pmatrix},  \\ \quad \begin{pmatrix}\textbf{l}_1\\\textbf{l}_2\\\textbf{l}_3\end{pmatrix}_{t}=D_{2} \begin{pmatrix}\textbf{l}_1\\\textbf{l}_2\\\textbf{l}_3\end{pmatrix}. \label{11.2}
\end{equation}
Here
\begin{eqnarray} 
C_{1}&=&\begin{pmatrix}0&\kappa_1&\sigma_{1} \\ -\kappa_1&0&\tau_{1} \\ -\sigma_{1}&-\tau_{1}&0 \end{pmatrix},  \quad
D_{1}=\begin{pmatrix}0&\omega_3&\omega_2 \\ -\omega_3&0&\omega_1 \\ -\omega_2 &-\omega_1&0 \end{pmatrix}, \label{11.3}\\ 
C_{2}&=&\begin{pmatrix}0&\kappa_{2}&\sigma_{2} \\ -\kappa_{2}&0&\tau_{2} \\ -\sigma_{2}&-\tau_{2}&0 \end{pmatrix},  \quad
D_{2}=\begin{pmatrix}0&\theta_3&\theta_2 \\ -\theta_3&0&\theta_1 \\ -\theta_2 &-\theta_1&0 \end{pmatrix},\label{11.4}
\end{eqnarray}
where  $\tau_{j}$ is the torsion, $k_j$ and  $\sigma_{j}$ are the geodesic curvature and normal curvature of the first and second curves, respectively. In Eq.(\ref{11.3})-(\ref{11.4}),  $\omega_j=\omega_{j}(k_{1},k_{2},\tau_{1}, \tau_{2},
\sigma_{1}, \sigma_{2}, ...)$   and $\theta_j=\theta_{j}(k_{1},k_{2},\tau_{1}, \tau_{2},
\sigma_{1}, \sigma_{2}, ...)$ are some real functions depending on $x$ and $t$. The compatibility condition of the linear equations (\ref{11.1})-(\ref{11.2}) are given by
\begin{eqnarray}  
C_{1t}-D_{1x}+[C_{1},D_{1}]&=&0, \\
C_{2t}-D_{2x}+[C_{2},D_{2}]&=&0.
\end{eqnarray}
In terms of elements these equations take the forms
\begin{eqnarray} 
\kappa_{1t}-\omega_{3x}-\sigma_{1}\omega_1+\tau_{1}\omega_2&=&0, \label{11.7}\\
\sigma_{1t}-\omega_{2x}+\kappa_1\omega_1-\tau_{1}\omega_3&=&0,  \label{11.8}\\
\tau_{1t}-\omega_{1x}-\kappa_1\omega_2+\sigma_{1}\omega_3&=&0 \label{11.9}
\end{eqnarray}
and
\begin{eqnarray}
\kappa_{2t}-\theta_{3x}-\sigma_2\theta_{1}+\tau_{2}\theta_{2}&=&0, \label{11.10}\\
\sigma_{2t}-\theta_{2x}+\kappa_{2}\theta_{1}-\tau_{2}\theta_{3}&=&0,  \label{11.11}\\
\tau_{2t}-\theta_{1x}-\kappa_{2}\theta_{2}+\sigma_2\theta_{3}&=&0. \label{11.12}
\end{eqnarray}
We now assume that $\sigma_{j}=\tau_{j}=0$. That is we consider the two interating plane curves. Then Eqs.(11.7)-(11.12) take the following simple forms
\begin{eqnarray} 
\kappa_{1t}-\omega_{3x}&=&0, \label{11.13}\\
\kappa_{2t}-\theta_{3x}&=&0. \label{11.14}
\end{eqnarray}
 In this paper, we assume that the  functions $\kappa_{j}, \theta_{3}, \omega_{3}$ have the following forms
\begin{equation} 
\kappa_1=q_{1t}-q_{1}^{2}+q_{1x}, \quad
\kappa_2=q_{2},  \quad
\theta_{3}=q_{2x}+2q_{1}q_{2}, \quad \omega_{3}=2q_{2x}, \label{11.15}
\end{equation} 
where $q_{j}=q_{j}(x,t)$ are some real functions. Hence and from Eqs.(11.1)-(11.2) we get the  the following Kairat-VII equation (K-VIIE):
\begin{eqnarray} 
\textbf{e}_{1tt}+\textbf{e}_{1tx}-\frac{\omega_{3t}+\omega_{3x}}{\kappa_{1}}
\textbf{e}_{1x}+(\omega_{3}^{2}+\omega_{3}\kappa_{1})
\textbf{e}_1&=&0,\\
\textbf{l}_{2tt}+\textbf{l}_{1tx}-\frac{\theta_{3t}+\theta_{3x}}{\kappa_{2}}
\textbf{l}_{1x}+(\theta_{3}^{2}+\theta_{3}\kappa_{2})
\textbf{l}_1&=&0,\\
q_{1tt}-2q_{1}q_{1t}+q_{1xt}-\left(\sqrt{\textbf{e}_{1x}^{2}}\right)_{t}&=&0, \\
q_{2t}-\left(\sqrt{\textbf{l}_{1x}^{2}}\right)_{t}&=&0.
\end{eqnarray}
Usually, this Kairat-VII equation we write in the following equivalent form
\begin{eqnarray} 
\textbf{A}_{tt}+\textbf{A}_{tx}-\frac{\omega_{3t}+\omega_{3x}}{\kappa_{1}}
\textbf{A}_{x}+(\omega_{3}^{2}+\omega_{3}\kappa_{1})
\textbf{A}&=&0,\\
\textbf{B}_{tt}+\textbf{B}_{tx}-\frac{\theta_{3t}+\theta_{3x}}{\kappa_{2}}
\textbf{B}_{x}+(\theta_{3}^{2}+\theta_{3}\kappa_{2})
\textbf{B}&=&0,\\
q_{1tt}-2q_{1}q_{1t}+q_{1xt}-\left(\sqrt{\textbf{A}_{x}^{2}}\right)_{t}&=&0, \\
q_{2t}-\left(\sqrt{\textbf{B}_{x}^{2}}\right)_{t}&=&0,
\end{eqnarray}
where $\textbf{A}\equiv\textbf{e}_{1}$ and $\textbf{B}\equiv\textbf{l}_{1}$ are the unit spin vectors that is $\textbf{A}^{2}=\textbf{B}^{2}=1$. This Kairat-VII equation is integrable. To prove this statement let us find the geometrical equivalent of the K-VIIE (11.18)-(11.19). To do that, let us plug the expressions (11.15) into the equations (11.13)-(11.14). As result,  we obtain the following Kairat-VIII equation (K-VIIIE) for the functions $q_{j}$:
\begin{eqnarray} 
q_{2t}-q_{2xx}-2(q_{1}q_{2})_{x}&=&0,\\
q_{1tt}+q_{1xxt}-(q_{1}^{2})_{xt}-2q_{2xx}&=&0. \label{3.10}
\end{eqnarray}
Thus the geometrical equivalent counterpart of the K-VIIE (11.18)-(11.19) is the K-VIIIE (11.20)-(11.21). The K-VIIIE is integrable. Its Lax representation is given by
\begin{eqnarray}
\phi_{xxx}+q_{1}\phi_{xx}-r\phi_{x}-(r_{x}+q_{1}r+q_{2})\phi&=&0,\\
\phi_{t}+\phi_{xx}-r\phi&=&0,
\end{eqnarray}
where $r_{t}=-2q_{2x}$. Finally, we note that the K-VIIIE can be written in the hamiltonian form. To do that, let us rewrite the K-VIIIE as
\begin{eqnarray} 
q_{2t}-q_{2xx}-2(q_{1}q_{2})_{x}&=&0,\\
q_{1t}+q_{1xx}-(q_{1}^{2})_{x}+r_{x}&=&0. \label{3.10}
\end{eqnarray}
 as 
 \begin{eqnarray} 
r_{1t}&=&\{r_{1},H\}= -\frac{\delta H}{\delta r_{2}}, \\
r_{2t}&=&\{r_{2}, H\}=\frac{\delta H}{\delta r_{1}}, \label{3.10}
\end{eqnarray}
where $r_{1}=q_{1}, \quad r_{2}=q_{2x}$, the Puasson bracket $\{,\}$  and the Hamilton function $H$  are given by
\begin{eqnarray} 
\{r_{1}(x,t),r_{2}(y,t)\}=-\delta(x-y), \label{3.10}
\end{eqnarray}
and
\begin{eqnarray} 
H=\int dx\left((r_{1}^{2}-r_{1x})r_{2x}+\frac{1}{2}r_{2xx}r_{2x}\right), \label{3.10}
\end{eqnarray}
respectively.
\section{Integrable motion of  plane  curves: K-IXE and K-XE}
In this section, we study the following Kairat-IX equation (K-IXE):
\begin{eqnarray} 
\textbf{A}_{tt}+\textbf{A}_{tx}-\frac{3((q_{x}q_{t})_{t}+(q_{x}q_{t})_{x})}{q_{t}+q_{xxx}}
\textbf{A}_{x}+3(q_{x}q_{t})(3q_{x}q_{t}+q_{t}+q_{xxx})
\textbf{A}&=&0,\\
q_{t}+q_{xx}-\sqrt{\textbf{A}_{x}^{2}}&=&0,
\end{eqnarray}
where $\textbf{A}(x,t)$ is the unit spin vector, $q(x,t)$ is a real function (potential). 
Let us now we will establish  the relation between the K-IXE and the differential geometry of plane curves. To do that, let us consider a plane  curve in the Euclidean space $R^{3}$. The corresponding  Serret - Frenet equation and its temporal  counterpart look like  
\begin{equation} 
\begin{pmatrix}\textbf{e}_1\\\textbf{e}_2\\\textbf{e}_3\end{pmatrix}_{x}=C \begin{pmatrix}\textbf{e}_1\\\textbf{e}_2\\\textbf{e}_3\end{pmatrix},  \\ \quad \begin{pmatrix}\textbf{e}_1\\\textbf{e}_2\\\textbf{e}_3\end{pmatrix}_{t}=D \begin{pmatrix}\textbf{e}_1\\\textbf{e}_2\\\textbf{e}_3\end{pmatrix}, \label{3.1}
\end{equation}
where $\textbf{e}_{3} =const$ that is $\textbf{e}_{3x}=\textbf{e}_{3t}=0$ and 
\begin{eqnarray} 
C&=&\begin{pmatrix}0&\kappa&0 \\ -\kappa&0&0 \\ 0&0&0 \end{pmatrix},  \quad
D=\begin{pmatrix}0&\omega&0 \\ -\omega&0&0 \\ 0 &0&0 \end{pmatrix}, \label{3.2}
\end{eqnarray}
where  $\kappa$ is the  curvature of the plane curve. The compatibility condition of the linear equation (\ref{3.1}) is  given by
\begin{eqnarray}  
C_{t}-D_{x}+[C,D]=0,
\end{eqnarray}
or in terms of elements 
\begin{eqnarray} 
\kappa_{t}-\omega_{x}=0. \label{12.6}
\end{eqnarray}
We now assume that $\kappa$ and $\omega$ have the forms
\begin{equation} 
\kappa=q_{t}+q_{xxx}, \quad
 \omega=3q_{x}q_{t}, \label{12.7}
\end{equation} 
We plug these expressions into  the equation (12.6) and, as result, we obtain  the following Kairat-X equation (K-XE):
\begin{eqnarray} 
q_{tt}+q_{xxxt}-3(q_{x}q_{t})_{x}=0. \label{12.8}
\end{eqnarray}
We can rewrite this K-XE as
\begin{eqnarray} 
q_{t}+q_{xxx}-p_{x}=0, \label{12.8}
\end{eqnarray}
where $p_{t}=3q_{x}q_{t}$.  The K-XE can be written in the hamiltonian form with   the Puasson bracket $\{,\}$  
\begin{eqnarray} 
\{r_{1}(x,t),r_{2}(y,t)\}=-\delta(x-y), \label{3.10}
\end{eqnarray}
and the Hamiltonian function
\begin{eqnarray} 
H=\int dx\left((r_{1}^{2}-r_{1x})r_{2x}+\frac{1}{2}r_{2xx}r_{2x}\right), \label{3.10}
\end{eqnarray}
respectively. Finally, we have  
 \begin{eqnarray} 
r_{1t}&=&\{r_{1},H\}= -\frac{\delta H}{\delta r_{2}}, \\
r_{2t}&=&\{r_{2}, H\}=\frac{\delta H}{\delta r_{1}}, \label{3.10}
\end{eqnarray}
This results tells us that the K-XE (12.8) is the geometrical equivalent counterpart of the K-IXE (12.1)-(12.2). Note that the K-XE (12.8) is integrable. The  Lax representation of the K-XE (12.8) is given by
\begin{eqnarray}
\phi_{xxxx}-3(q_{x}\phi_{x})_{x}+q_{t}\phi&=&0,\\
\phi_{t}+\phi_{xxx}-3q_{x}\phi&=&0.
\end{eqnarray}
The compatibility condition of these linear equations $\phi_{xxxxt}=\phi_{txxxx}$ gives the K-XE (12.8). 

\section{The Kairat-Kuralay-Myrzakulov-Shynaray equation}

In this section, we consider the following Kairat-Kuralay-Myrzakulov-Shynaray equation (KKMSE):
\begin{eqnarray}
q_{t}-\frac{1}{b}uq_{x}+\frac{2\beta}{b} qq_{t}-\beta r_{t}&=&0, \label{13.1} \\
r_{t}-\frac{1}{b}ur_{x}+\frac{2\beta}{b} rq_{t}-\frac{\beta}{2ab}q_{xxt}&=&0,\label{13.2}\\
u_{x}+\beta q_{t}&=&0, \label{13.3}
\end{eqnarray}
where $a,b, \beta$ are real constants. If we assume that  $a=\beta=1, \quad b=2$, then the  KKMSE  (\ref{13.1})-(\ref{13.3}) takes the form
\begin{eqnarray}
q_{t}-\frac{1}{2}uq_{x}+ qq_{t}- r_{t}&=&0, \label{13.4} \\
r_{t}-\frac{1}{2}ur_{x}+ rq_{t}-\frac{1}{4}q_{xxt}&=&0,\label{13.5}\\
u_{x}+ q_{t}&=&0, \label{13.6}
\end{eqnarray}
 We note that the  KKMSE  (13.1)-(13.3) is integrable. Its   Lax representation   is given by
\begin{eqnarray}
\Phi_{x}&=&U_{13}\Phi, \label{13.7}\\
\Phi_{t}&=&V_{13}\Phi, \label{13.8}
\end{eqnarray}
where
\begin{eqnarray}
U_{13}&=&\begin{pmatrix}0&a\\ b\lambda^{2}+q\lambda+r&0  \end{pmatrix},   \quad
V_{13}=\frac{1}{1-\beta \lambda}B, \quad B=B_{2}\lambda^{2}+B_{1}\lambda +B_{0}, \\
B_{2}&=&\begin{pmatrix}0& 0 \\ u & 0
\end{pmatrix}, \quad B_{1}=\begin{pmatrix}0& 0 \\ b^{-1}uq & 0
\end{pmatrix}, \\
B_{0}&=&\begin{pmatrix}\frac{\beta }{2b}q_{t}& ab^{-1}u \\b^{-1}ur+\frac{\beta}{2ab}q_{xt}& -\frac{\beta }{2b}q_{t}
\end{pmatrix}. 
\end{eqnarray}
The compatibility condition $\Phi_{xt}=\Phi_{tx}$  of the set of linear equations (\ref{13.7})-(\ref{13.8}) that is   
\begin{eqnarray}
U_{13t}-V_{13x}+[U_{13},V_{13}]=0 \label{13.12}
\end{eqnarray}
gives the Kairat-Kuralay-Myrzakulov-Shynaray equation (KKMSE)  (13.1)-(13.3). Finally we note that as the  integrable equation, the KKMSE (\ref{13.1})-(\ref{13.3}) admits the  N-soliton solution, infinite number of conservation laws, Hamiltonian structure and so on. 

\section{Zhanbota transcendents}

 There exist some integrable Zhanbota equations in 1+1 and 1+0 dimensions: Zhanbota-I equation, the Zhanbota-II equation, the Zhanbota-III equation and so on.  In this section, we want to present some of them. 

\subsection{Zhanbota-III equation}

One of interesting integrable spin systems in 1+0 dimensions is the Zhanbota-III (Z-III) equation. It has the form
\begin{eqnarray}
S_{xx}+\frac{1}{2}tr(S_{x}^{2})S-\frac{1}{2}(\ln(tr(S_{x}^{2})))_{x}S_{x}+(\beta+\epsilon\sqrt{tr(S_{x}^{2}})[S,S_{x}]=0, \label{14.1}
\end{eqnarray}
where $\epsilon=\pm 1$ and $\beta$ is some complex constant. This Z-III  equation  is integrable. The corresponding Lax representation is given by
\begin{eqnarray}
\Psi_{x}&=&U_{III}\Psi, \label{14.2}\\
\Psi_{\lambda}&=&V_{III}\Psi, \label{14.3}
\end{eqnarray}
where
\begin{eqnarray}
U_{III}&=&\left(\lambda-\beta+\frac{q_{III}}{\lambda}-\frac{q_{III}}{\beta}\right)S+\frac{\beta}{4}\left(\frac{1}{\lambda}-\frac{1}{\beta}\right)[S,S_{x}], \label{14.4}\\
V_{III}&=&(4\lambda^{4}+2q_{III}^{2}+x)S+\frac{\beta}{2q_{III}}\left(2\lambda q_{IIIx}+\frac{1}{2\lambda}\right)S_{x}+\frac{\beta}{4q_{III}}\left(4\lambda^{2}q_{III}+2q_{III}^{2}+x\right)[S,S_{x}],\label{14.5}
\end{eqnarray}
with  
\begin{eqnarray}
q_{III}=\pm 0.5\beta\sqrt{{\bf S}_{x}^{2}}, \quad {\bf S}_{x}^{2}=0.5tr(S_{x}^{2}), \quad q_{III}^{2}=0.25\beta^{2}{\bf S}_{x}^{2}=\frac{\beta^{2}}{8}tr(S_{x}^{2}).\label{14.6}
\end{eqnarray}
The compatibility condition of the linear equations (\ref{14.2})-(\ref{14.3}) $\Psi_{x\lambda}=\Psi_{\lambda x}$ gives
\begin{eqnarray}
U_{III\lambda}-V_{IIIx}+[U_{III},V_{III}]=0\label{14.7}
\end{eqnarray}
which  is equivalent to the Z-III equation (13.1). 
\subsection{Zhanbota-IV equation}
Another example of integrable spin systems in 1+0 dimensions is the Zhanbota-IV  (Z-IV) equation. The Z-IV equation   reads as 
\begin{eqnarray}
S_{xx}+\frac{1}{2}tr(S_{x}^{2})S-\frac{1}{2}\left(\ln\left(-\frac{tr(S_{x}^{2})}{8}\right)\right)_{x}S_{x}-i\beta[S,S_{x}]=0, \label{14.8}
\end{eqnarray}
where $\beta$ is an complex constant. As integrable equation, this Z-IV equation  admits the Lax representation, infinite number of conservation  laws and so on.  For example, its  Lax representation has the form
\begin{eqnarray}
\Psi_{x}&=&U_{IV}\Psi, \\
\Psi_{\lambda}&=&V_{IV}\Psi, \label{14.10}
\end{eqnarray}
where
\begin{eqnarray}
U_{IV}&=&-i(\lambda-\beta)S, \label{13.11}\\
V_{IV}&=&-i(4\lambda^{2}+2q_{IV}^{2}+x)S+\frac{iq_{IVx}}{q_{IV}}S_{x}+\left(\lambda-\frac{\alpha}{4\lambda q_{IV}}\right)[S,S_{x}].
\end{eqnarray}
In this case  
\begin{eqnarray}
q_{IV}=\pm i 0.25\sqrt{{\bf S}_{x}^{2}}, \quad {\bf S}_{x}^{2}=0.5tr(S_{x}^{2}), \quad q_{IV}^{2}=-0.25{\bf S}_{x}^{2}=-\frac{1}{8}tr(S_{x}^{2}).\label{14.13}
\end{eqnarray}
From the compatibility condition of the linear equations (13.9)-(13.10) $\Psi_{x\lambda}=\Psi_{\lambda x}$:
\begin{eqnarray}
U_{IV\lambda}-V_{IVx}+[U_{IV},V_{IV}]=0\label{14.14}
\end{eqnarray}
we obtain  the Z-IV equation (13.8). 

\section{Conclusion} 

The goal of this paper is to attract more attention to some Kairat equations.  These Kairat equations are integrable  are intimately related to the very known  equations. 
In the present paper, we provided a geometrical  description some of integrable Kairat equations. It is done in two
cases: geometrical and gauge equivalences between these equations. We studied the Lax integrability of these  nonlinear evolutionary equations. In particular, we explicitly constructed their Lax representations. For the Kairat-II equation the three travelling wave solutions, namely, elliptic, solitonic and rathional solutions  are found. We also presented two integrals of motion of  this equation. The techniques used in this paper can be adopted to study  other integrable nonlinear  equations (see, e.g., \cite{51}-\cite{56} and references therein). Finally, we consider some integrable Zhanbota equations.

\section*{Acknowledgments}
This work was supported by the Ministry of Science and Higher Education of the Republic of Kazakhstan, Grant AP14870191.

\end{document}